\documentclass[%
	aps,prb,10pt,twocolumn,raggedbottom,nobalancelastpage]%
	{revtex4-1}
\usepackage{graphicx}
\usepackage{amsmath}
\usepackage{braket}

\newcommand{\dvec}[1]{\ensuremath{\boldsymbol{#1}}}
\newcommand{\vk}{\dvec{\mathrm{k}}}
\newcommand{\vq}{\dvec{\mathrm{q}}}
\renewcommand{\vr}{\dvec{\mathrm{r}}}

\newcommand{\nimp}{\ensuremath{n_{\mathrm{i}}}}
\newcommand{\cmsq}{\ensuremath{\mathrm{cm}^{-2}}}
\newcommand{\meV}{\ensuremath{\mathrm{meV}}}
\newcommand{\EF}{\ensuremath{E_{\mathrm{F}}}}
\newcommand{\vF}{\ensuremath{v_{\mathrm{F}}}}
\newcommand{\qs}{\ensuremath{q_\mathrm{s}}}
\newcommand{\dsc}{\ensuremath{d_\mathrm{sc}}}

\newcommand{\qTF}{\ensuremath{q_{\mathrm{TF}}}}

\begin{document}
\title{Comparison of microscopic models for disorder in bilayer
graphene: Implications for the density of states and the optical
conductivity}
\author{D. S. L. Abergel}
\author{Hongki Min}
\author{E. H. Hwang}
\author{S. Das Sarma}
\affiliation{Condensed Matter Theory Center, Department of Physics,
University of Maryland, College Park, MD 20742, USA}

\begin{abstract}
We study the effects of disorder on bilayer graphene using four
different microscopic models and directly compare their results. We
compute the self-energy, density of states, and optical conductivity
in the presence of short-ranged scatterers and screened Coulomb
impurities, using both the Born approximation and self-consistent Born
approximation for the self-energy. We also include a finite interlayer
potential asymmetry which generates a gap between the
valence and conduction bands. We find that the qualitative behavior
of the two scattering potentials are similar, but that the choice of
approximation for the self-energy leads to important differences near
the band edge in the gapped case. Finally, we describe how these
differences manifest in the measurement of the band gap in optical and
transport experimental techniques.
\end{abstract}

\maketitle

\section{Introduction \label{sec:intro}}

The current interest in bilayer graphene\cite{dassarma-rmp2011,
abergel-advphys2010} is motivated largely by the possibility of its
application in the designing of novel electronic devices, as well as the
opportunity to investigate the fundamental physics of massive chiral
electrons in a condensed matter setting.
In particular, the opportunity to open a dynamically tunable
band gap by electronic gating was predicted theoretically
\cite{mccann-prb2006}, and verified in optical experiments
\cite{ohta-sci2006, zhang-nature2009, mak-prl2009, kuzmenko-prb2009,
li-prl2009}, and this presents an opportunity not currently available in
traditional two-dimensional electron systems\cite{ando-rmp1982}.
However, transport measurements 
\cite{oostinga-natmat2007, zou-prb2010, xiao-prb2010, yan-nl2010,
xia-nl2010, taychatanapat-prl2010, jing-nl2010}
demonstrate that the full nature of the effects of disorder induced in
the graphene by its environment\cite{mucciolo-jpcm22} is not currently
resolved. 
This is a crucial issue since it appears that environmental disorder is
the main limitation on the favorable properties of bilayer graphene
devices.
In particular, there is a wide discrepancy in the size of the band gap
extracted from transport measurements (which has obvious negative impact
on the application of bilayer graphene in switching devices)
and significant variation in the sub-gap conductivity as a function of
temperature.
At low temperature, variable range hopping via midgap states dominates
and produces a relatively small value for the gap, while at higher
temperature, thermally activated transport between the band tails is the
predominant mechanism\cite{yan-nl2010}.
It has been suggested \cite{dassarma-prb2010, hwang-prb2010} that both
charged impurity disorder and short-range defect scattering play a role
in transport properties of bilayer graphene and the puddling of
electrons due to charged impurity disorder was also shown to be crucial
in the understanding of capacitance measurements of dual-gated bilayer
graphene\cite{abergel-prb2011, young-arXiv, henriksen-prb2010}.
A very recent work \cite{rossi-arXiv} shows that the percolation
transport gap and the spectral band gap could be significantly different
in bilayer graphene due to the strong potential fluctuation and density
inhomogeneity induced by electron-hole puddles arising from random
charged impurities in the environment.
Therefore, disorder is a key effect in many different measurements and a
thorough understanding of the microscopic origin and effects of
impurities and other scatterers is highly important.

Previously, theoretical study of the role of disorder in bilayer
graphene has been undertaken
\cite{koshino-prb2006, bena-prl2008, nilsson-prb2008, ando-jpsj2011,
ferreira-prb2011, yuan-prb2010, castro-prl2010, mkhitaryan-prb2008,
dahal-prb2008, nilsson-prl2007} in which several types of disorder (such
as resonant scatterers, lattice vacancies,
short-range scatterers, and generic scattering processes) were
investigated within a variety of theoretical schemes (such as the
coherent potential approximation, the Born approximation and
self-consistent Born approximation, as well as other numerical
techniques). The effect of disorder on the density of states (DOS), 
dc and ac conductivities, and other physical quantities were studied.
However, to date, there has not been a systematic comparison of the most
physical scattering mechanisms -- charged impurity disorder and
short-range scatterers -- for different approximation schemes. This is
the topic of the current article.
In this paper, we present a comprehensive analysis of the effect of
charged impurity disorder and short-range scatterers on the DOS
and optical conductivity of bilayer graphene in the presence of a
band gap created by electrostatic gating. 
We compare these two scattering mechanisms which model the
electron-impurity interaction, and two approximations to the self-energy
induced by this interaction in the perturbation theory. Specifically, we
consider the screened electron--impurity Coulomb interaction (CI) and
the short-ranged interaction (SRI) modeled as a $\delta$-function
potential in real space as the model disorder potentials for our study. 
This paper is the first analysis which treats these two scattering
mechanisms on an equal footing and makes direct comparisons between
them, since previous works have treated only different forms of
short-ranged electron--impurity interactions, and we extend our analysis
of the CI presented previously\cite{min-prb2011}.
For both of the these electron--impurity interaction potentials, we
describe the Born approximation (BA) and self-consistent Born
approximation (SCBA) and highlight the differences between these two
levels of approximation in the perturbation theory. 
We also consider both the gapped and gapless situations.
It should be noted that some of the numerical techniques which were
previously applied to the electron--impurity interaction
\cite{ferreira-prb2011, yuan-prb2010} technically go beyond the BA and
SCBA and their application to the gapped case would be desirable.
However, we believe that particularly the SCBA will capture the relevant
physics near the band edge.

Although our main interest is a comprehensive theoretical formal
understanding of disorder effects on bilayer electronic properties using
a number of different models, a strong motivation for our work is the
large discrepancy between the experimentally extracted transport gap in
bilayer graphene and the theoretically calculated band gap obtained from
band structure calculations. We also want to understand the effect of
disorder on the optical conductivity of bilayer graphene since
extracting an optical band gap from the optical conductivity data is
non-trivial in the presence of strong disorder (although it is routinely
done in the literature, often incorrectly in our view). Our work should
have implications for the bilayer density of states (and hence the
transport gap) and the optical gap extracted from optical measurements.
We also mention a particular advantageous feature of our theory is that
we use the accurate four band model for bilayer graphene throughout this
work and retain the contributions from all four bands, avoiding various
simplifications used in the theoretical literature dealing with bilayer
graphene.

The structure of this paper is as follows. In Sec.~\ref{sec:Sigma} we
describe the self-energy of electrons in bilayer graphene using the
short-range and Coulomb scattering potentials for the BA and SCBA, 
and in Sec.~\ref{sec:DOS} we apply the resulting Green's functions to
compute the DOS.
In Sec.~\ref{sec:optcond} we describe the optical conductivity and then 
in Sec.~\ref{sec:gaps} we focus on the highly important issue of
the band gap and how it is extracted from various experimental
techniques and describe the role of disorder in that context.
Finally, we summarize our results in Sec.~\ref{sec:summary} and
highlight the most important conclusions of our work.
Two appendices contain additional results: A scheme to use the
computationally efficient SRI to approximate the CI for low carrier
density, and the derivation of the optical conductivity.

\begin{figure}[tbp]
	\includegraphics[]{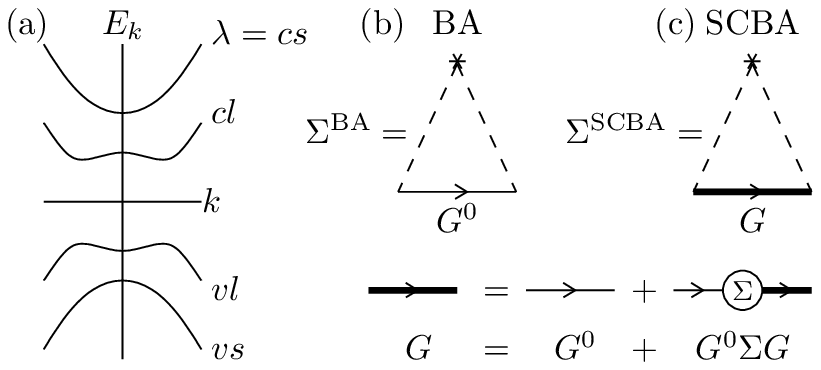}
	\includegraphics[]{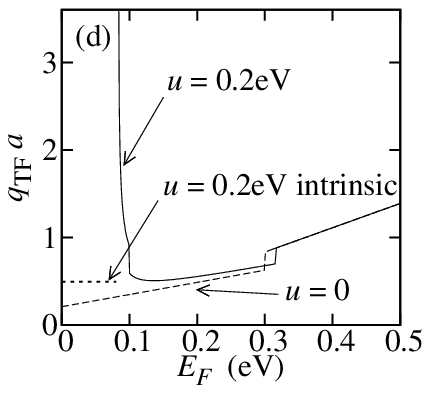}
	\includegraphics[]{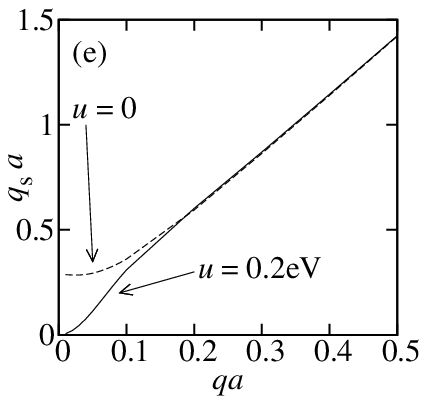}
	\caption{(a) Sketch of the low-energy band structure of bilayer
	graphene in the presence of an external electric field which creates
	a band gap at low energy. 
	The bands are labeled with the index $c,v$ for conduction and
	valence bands, and $l,s$ for low-energy and split bands
	respectively.
	Diagrams corresponding to (b) the Born approximation and
	(c) the self-consistent Born approximation for the self-energy, and
	the definition of the Green's function in terms of the self-energy.
	(d) The Thomas-Fermi screening wave vector $\qTF$ as a function of
	the Fermi energy. The band edge in the $u=0.2\mathrm{eV}$ case is at
	$\EF = 0.083\mathrm{eV}$.
	(e) The screening wave vector in the RPA as a function of wave
	vector in the intrinsic case.
	\label{fig:diagrams}}
\end{figure}

We now introduce the single-particle theory which underlies our
considerations of disorder. Throughout this paper, we use the four band
tight-binding model for bilayer graphene but neglect next-nearest
neighbor hoppings between the two layers.
We label the electron states with the band index $c,v$ for the
conduction and valence bands, and with the branch index $l,s$ for the
low-energy and split bands. The dispersion of the four bands is given by
the equation\cite{abergel-advphys2010, mccann-prb2006}
\begin{equation}
	E_{\vk\lambda} = \nu_\lambda \sqrt{ \zeta^2 +
	\tfrac{\gamma_1^2}{2} + \tfrac{u^2}{4} + b_\lambda \sqrt{
	\tfrac{\gamma_1^4}{4} + \zeta^2 ( u^2 + \gamma_1^2 ) }}
\end{equation}
where $\lambda=\{ \nu_\lambda, b_\lambda\}$ labels the band, 
$\nu=+1(-1)$ for the conduction (valence) band, and $b=+1(-1)$ for
the split (low-energy) branch [see Fig. \ref{fig:diagrams}(a)],
and $\zeta = \hbar \vF k$. The inter-layer coupling in the tight
binding formalism is $\gamma_1$, and $\vF$ is the Fermi velocity of
monolayer graphene\cite{abergel-advphys2010,dassarma-rmp2011}. 
This band structure is isotropic in the wave vector and symmetric
between conduction and valence bands.
In the presence of an external electric field perpendicular to the plane
of the graphene, an asymmetry is introduced in the potential on the
upper and lower layers which generates a gap between the $cl$ and $vl$
bands at low energy.
This gap is parameterized by the energy $u$, corresponding to the energy
difference between the two bands at $k=0$. However, the minimum of the
band gap is found at a wave vector $k_\mathrm{gap} =
\frac{u}{2\hbar\vF}\sqrt{\frac{2\gamma_1^2+u^2}{\gamma_1^2+u^2}}$ 
and is given by $E_\mathrm{gap}^0 = u\gamma_1 / \sqrt{u^2+\gamma_1^2}$.
This is known as the ``sombrero'' shape.
We do not include various next-nearest neighbor interlayer hops which
have been considered elsewhere\cite{dassarma-rmp2011,
abergel-advphys2010}. 
We justify this by noting that, for example, the energy of the
inter-band optical transitions near the K point are only modified by a
few percent of the interlayer coupling $\gamma_1$ when these hops are
included. In a similar way, while there will be some small quantitative
difference in the DOS due to inclusion of these higher order terms, the
position of the band edge is almost unchanged and the disorder is by far
the dominant effect. Therefore the simple tight-binding model we outline
is sufficient to obtain an accurate description of the DOS and optical
conductivity in the presence of disorder.

The two-dimensional CI in momentum space is
\begin{equation}
	V^{\mathrm{C}}(\vq) = \frac{2\pi e^2}{\kappa (q+\qs)} e^{-qd}
	\label{eq:VCdef}
\end{equation}
where $d$ is the distance of the impurities from the graphene surface,
$\kappa$ is the effective dielectric constant of the system, and $\qs$
is the screening wave vector, which can in principle be a function of
$q$.
In the extrinsic case (\textit{i.e.} when there is a finite density of
electrons or holes in the graphene so that $E_F\neq 0$), 
the screening is taken into account in the Thomas-Fermi
approximation\cite{hwang-prb2007} so that the $q$ dependence of the
screening wave vector is constant
and
\begin{equation}
	\qTF = \frac{2\pi e^2}{\kappa} D_0(\EF)
	\label{eq:qTFdef}
\end{equation}
where $D_0(\EF)$ is the non-disordered D OS at the Fermi energy.
This screening wave vector is shown in Fig.~\ref{fig:diagrams}(d) as a
function of the Fermi energy. In the ungapped case (dashed line), the
DOS is finite for all values of $\EF$ so $\qTF$ is well defined. The
shape of the DOS is replicated in $\qTF$ so that it increases linearly
except for the onset of the split bands which causes a step in the DOS
at $E_F \approx 0.3\mathrm{eV}$.
Note that this linear increase is a result of the more accurate
hyperbolic model which we use for the dispersion of bilayer graphene
near the K point. The quadratic approximation is only valid at low
density and predicts a constant DOS, however the density ranges we
consider exceed the applicability of this model\cite{hwang-prl2008}.
When the gap is present, the DOS is a more complex function which is
reflected in $\qTF$ [solid line in Fig.~\ref{fig:diagrams}(d)]. 
In particular, it is divergent at the band edge
so that the screening becomes very strong for low carrier density.
For intrinsic bilayer graphene (that is, when the carrier density is
zero so that $E_F=0$) in the gapped regime, the density of states is not
defined and some other approximation to the screening wave vector must
be made to avoid the unphysical divergence of the Coulomb interaction at
$q=0$.
In Fig.~\ref{fig:diagrams}(e), we show the screening wave vector as a
function of $q$ in the random phase approximation (RPA) for intrinsic
bilayer graphene. 
This shows that for $u=0$ the $q=0$ value is finite, which is consistent
with an earlier analysis of the screening in bilayer
graphene\cite{hwang-prl2008} computed using a quadratic approximation
for the band structure. 
For $u=0.2\mathrm{eV}$, the low-$q$ behavior is almost linear with the 
$q = 0$ value being zero. 
In both cases, the screening wave vector depends linearly on $q$ at
high wave vector.
Therefore, the $q\to 0$ limit shows that this more sophisticated
approximation also fails to remove the divergence in $V^C(\vq)$ for
gapped bilayer graphene.
In order to make an order-of-magnitude estimate for the screening in
this case, we take the extrapolation of the higher-energy screening wave
vector to $q=0$ [as shown by the dotted line in 
Fig.~\ref{fig:diagrams}(d)].
For $u=0.2\mathrm{eV}$, this means $\qTF a \approx 0.5$ where
$a=0.246\mathrm{nm}$ is the lattice constant.
The precise value chosen will not significantly affect the DOS or
optical conductivity.
The regularization of the CI at $q=0$ is necessary for avoiding and
artificial divergence, but the details of this infrared regularization
do not affect the results presented in this paper.

The short-range interaction (SRI) is given by the potential
\begin{equation}
	V^{\mathrm{SR}}(\vr) = V_0 \delta(\vr) 
	\quad\Rightarrow\quad V^{\mathrm{SR}}(\vq) = V_0
	\label{eq:VSRIdef}
\end{equation}
where the interaction strength is parameterized by $V_0$. In principle,
the parameter $V_0$ is not known so it can either be used as a fitting
parameter, or some physical reasoning can be used to estimate its value.
We make a phenomenological assumption that $V_0 = 2\pi e^2
\dsc/\kappa$ where $\dsc$ is the scattering
cross-section of the short-ranged impurities. We assume that $\dsc =
1\mathrm{nm}$ and that the sample is mounted on an SiO$_2$ substrate so
that $\kappa=2.5$ is the effective dielectric constant of the
environment, giving $V_0 = 3.62 \mathrm{eV} \mathrm{nm}^2$.
In Appendix \ref{app:SRCIcomp} we demonstrate a way in which the SRI can
be used to approximate the CI in the low-energy region.
Our models of long- and zero-ranged disorder give similar results at low
energies (see Appendix \ref{app:SRCIcomp}) so our model also applies in
this energy range to the intermediate-range resonant scattering
disorder. In the literature, it is noted that the resonant scatterers
are likely to also contribute to the existence of mid-gap states in the
system which, as far as we know, have never been observed in any optical
experiments. We do not take these mid-gap states into account in our
model, but mention that it should be straightforward to include such
states in our theory if experiments demonstrate their existence.

\section{Self-energy} \label{sec:Sigma}

\begin{figure*}[tbp]
	\centering
	\includegraphics[]{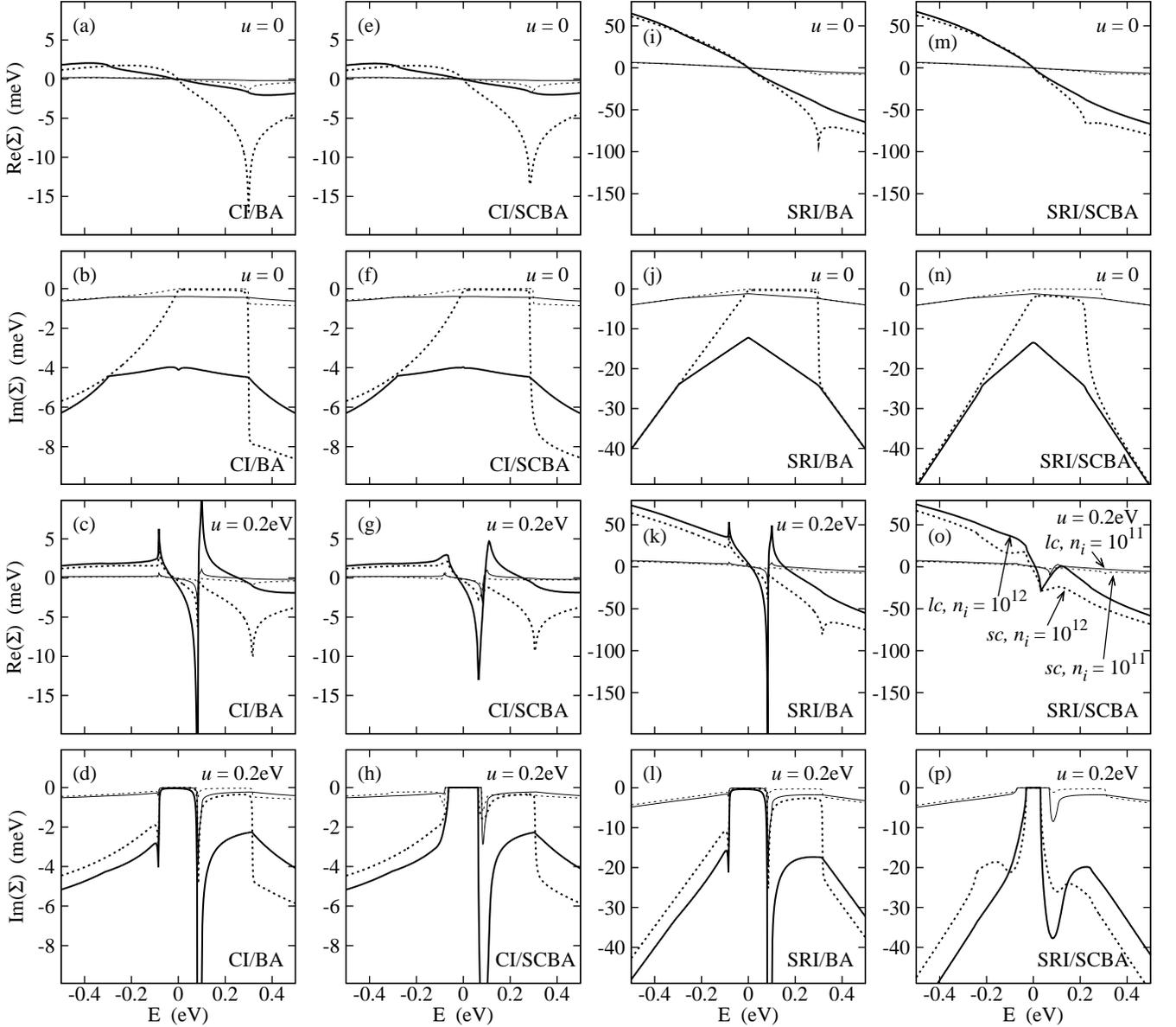}
	\caption{The real and imaginary parts of the self-energy in the
	low-energy conduction band (solid lines) and split conduction band
	(dashed) lines for $\nimp = 10^{11}\cmsq$ (red lines) and $\nimp =
	10^{12}\cmsq$ (black lines) at $|\vk|a=0.01$.  
	The scattering potential (CI or SRI), approxmation for the
	self-energy (BA or SCBA) and band gap ($u=0$ or $u=0.2\mathrm{eV}$)
	are labeled in each panel. 
	The self-energy is trivially zero in the clean limit.
	\label{fig:Sigma}}
\end{figure*}

In this section, we describe the self-energy introduced by interaction
between the electrons and the impurities in bilayer graphene. We use two
approximations, the BA and the SCBA, and compare the results of each.
Formally, the self-energy in the BA is
\begin{equation}
	\Sigma_{\vk\lambda}^{\mathrm{BA}}(E) = \nimp \sum_{\vk'\lambda'}
	\frac{|V(\vk-\vk')|^2 F_{\vk\lambda,\vk'\lambda'}}
		{E - E_{\vk'\lambda'} + i\eta}
	\label{eq:defBA}
\end{equation}
and in the SCBA is
\begin{equation}
	\Sigma_{\vk\lambda}^\mathrm{SCBA}(E) = \nimp \sum_{\vk'\lambda'}
	\frac{|V(\vk-\vk')|^2 F_{\vk\lambda,\vk'\lambda'}}
		{E - E_{\vk'\lambda'} - \Sigma^{\mathrm{SCBA}}_{\vk'\lambda'}(E)}
	\label{eq:defSCBA}
\end{equation}
which differs from the BA because it includes the full Green's function
on the right-hand side and therefore defines a self-consistent equation
for the self-energy. The SCBA self-energy is generally found by iterating
this expression taking the BA as the initial value of the self-energy on
the right-hand side until convergence is reached. 
The diagrams corresponding to these approximations are shown in Fig.
\ref{fig:diagrams}(b,c).
Although the SCBA is a self-consistent approximation involving an
infinite series of diagrams, it is not always necessarily better that
the BA since in each order many diagrams (e.g. the crossing diagrams)
are left out -- in general, the self-consistency in
the SCBA tends to smooth out the sharp features in the BA results as we
shall see below in our calculations of the DOS.
In these equations, the symbol $\nimp$ is the areal density of charged
impurities, $V(\vk)$ is the electron-impurity interaction potential for
the wave vector $\vk$, $F_{\vk\lambda,\vk'\lambda'}$ is the wave
function overlap between $\vk\lambda$ and $\vk'\lambda'$ states, and
$\eta$ is a positive infinitesimal.
In the following we shall describe the self-energy of both the CI and
the SRI each of which requires the appropriate expression for the
interaction potential [Eq.~\eqref{eq:VCdef} or Eq.~\eqref{eq:VSRIdef}]
to be substituted into the equations for the self-energy.
Because the bands are electron-hole symmetric, we also have the symmetry
relation
\begin{equation}
	\Sigma_{\vk c\nu}(E) = - \Sigma_{\vk v\nu}^{\ast}(-E).
	\label{eq:SigmaSymmetry}
\end{equation}
The real and imaginary parts of these self-energies are plotted in
Fig~\ref{fig:Sigma}. 
Specifically, we take wave vector $|\vk|a=0.01$, impurity concentrations
$\nimp = 10^{11}\cmsq$ and $\nimp=10^{12}\cmsq$, $\kappa =
2.5$ corresponding to a back-gated SiO$_2$ sample,
$\eta=10^{-3}\mathrm{eV}$
and two values of the gap: $u=0$ and $u=0.2\mathrm{eV}$ as indicated by
the labels in
Fig.~\ref{fig:Sigma}. The solid lines refer to the $lc$ band, the dashed
lines to the $sc$ band. The $lv$ and $sv$ bands can be found via
Eq.~\eqref{eq:SigmaSymmetry}. In the CI case, we have taken $n=5\times
10^{12}\cmsq$ to define the screening wave vector.
Throughout this article, we have $\gamma_1=0.30\mathrm{eV}$, $\gamma_0 =
3\mathrm{eV}$ which gives $\vF = \sqrt{3}\gamma_0 a/(2\hbar) = 9.75\times
10^5 \mathrm{ms}^{-1}$, and $d=0$.
In the SRI/BA, the factor $\nimp V_0^2$ is a multiplier, meaning that
the lines for the different impurity concentrations are scaled copies of
each other, but this is not the case for the CI or for the SCBA.
We see that the BA and SCBA give rather similar
results for most values of energy, especially in the gapless case. 
However, the sharp spikes in the BA due to band edges are rounded off in
the SCBA, and the onset of the imaginary part at the band
edge is shifted to slightly lower energy in the SCBA.
These differences will be manifest in the DOS at the band edge as we
discuss below. In the $u=0$ case, the imaginary part of the low-energy
branch is always finite, and is roughly constant up to the energy where
the split bands become occupied. 
When $u$ is finite, the imaginary part disappears in the region near
$E=0$ signifying that a gap has opened. Large peaks appear near the band
edge, and these peaks are not symmetrical in energy, but are larger on
the positive energy (conduction band) side of the gap. This behavior is
reversed in the valence band due to the relation in
Eq.~\eqref{eq:SigmaSymmetry}. 
The presence of the gap also modifies the inter-band contribution to the
self-energy, reducing the size of the imaginary part in the higher
energy regions. 
In addition to the change in the relative size of the self-energy,
increasing impurity concentration has the effect of introducing a larger
shift to the energy at which the band minima occur.

\section{Density of states} \label{sec:DOS}

\begin{figure*}[tbp]
	\centering
	\includegraphics[]{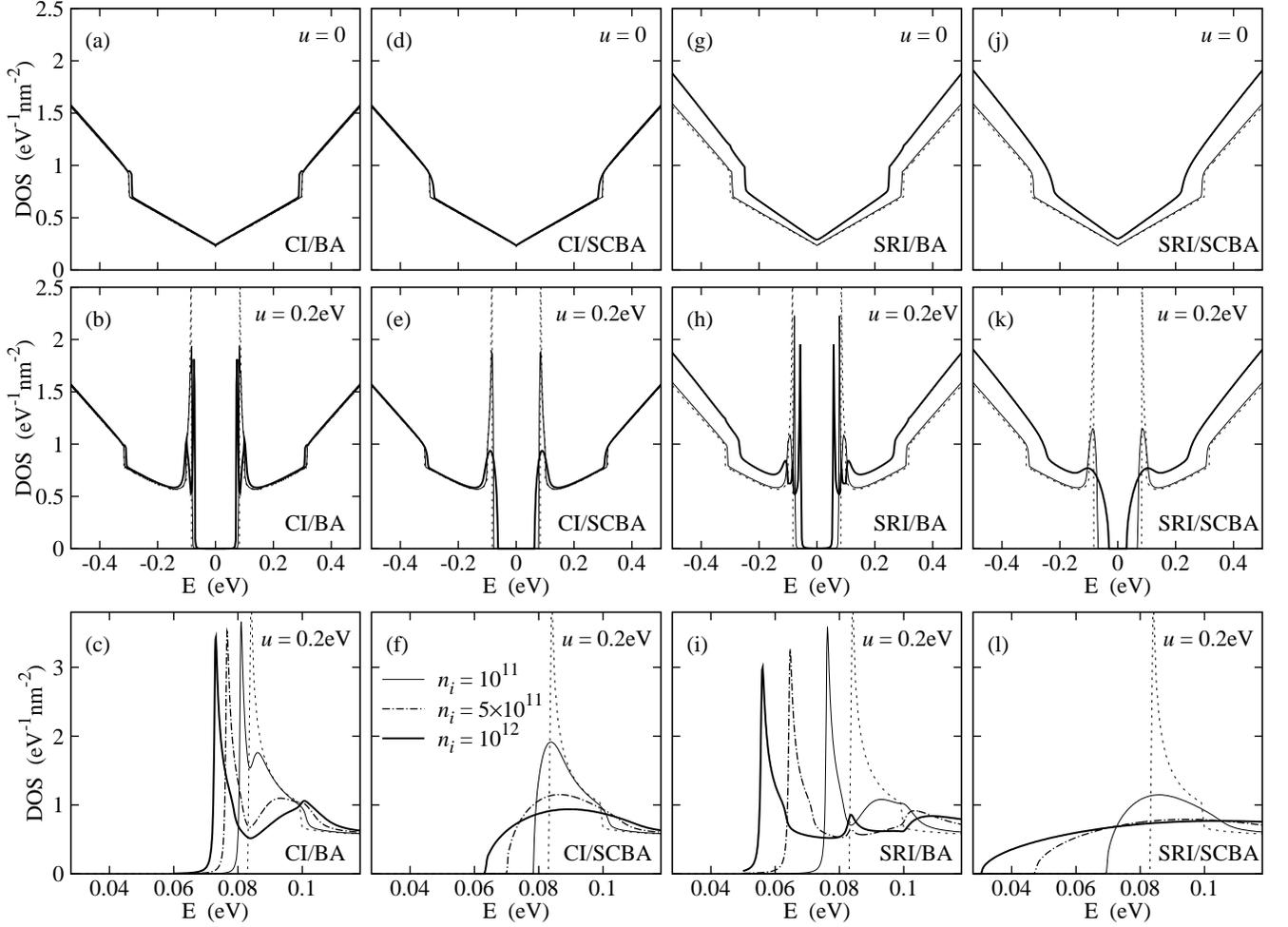}
	\caption{The density of states for $\nimp=10^{11}\cmsq$ (red lines),
	$\nimp = 10^{12}\cmsq$ (black lines), and (near the band edge) for
	$\nimp=5\times 10^{11}\cmsq$ (blue lines). The scattering potential
	(CI or SRI), approximation for the self-energy (BA or SCBA), and gap
	($u=0$ or $u=0.2\mathrm{eV}$) are labeled in each panel. The dotted
	lines give the DOS in the clean limit.
	\label{fig:DOS}}
\end{figure*}

These self-energies described in Section \ref{sec:Sigma} define Green's 
functions which can be used to compute the DOS as 
\begin{equation}
	D(E) = -\frac{g_\mathrm{s} g_\mathrm{v}}{\pi} \sum_{\vk\lambda} 
	\mathrm{Im} G_{\vk\lambda}(E)
\end{equation}
where $g_\mathrm{s}$ and $g_\mathrm{v}$ are the spin and valley
degeneracies respectively, and 
$G_{\vk\lambda}(E) = \left[ E - E_{\vk\lambda} -
\Sigma_{\vk\lambda}(E) \right]^{-1}$ is the Green's function
incorporating the self-energy induced by the electron-impurity
interaction (see Fig.~\ref{fig:diagrams}). 
The DOS for the clean system is found by substituting
$\Sigma_{\vk \lambda}(E) = -i\eta$ (where $\eta$ is a positive 
infinitesimal) or by exact extraction from the band structure.
The DOS functions are shown in Fig.~\ref{fig:DOS} for the BA and the
SCBA, for the CI and SRI, at two different values of impurity
concentration, and for the ungapped and gapped cases.
In the gapless situation (top row of Fig.~\ref{fig:DOS}) the only
difference between the two approximations is a slight rounding-off of
the step at the onset of the split bands in the SCBA. For our choice of
parameters, the SRI gives a stronger change in the slope of the DOS with
increasing impurity density than the CI, leading to a larger change in
the DOS at higher energies.
Also, increasing impurity concentration shifts the bottom of the split
bands to lower energy, as indicated by the position of the step-like
feature. 
In the gapped case (middle row of Fig.~\ref{fig:DOS}), increasing
impurity concentration has the marked effect of lowering the energy of
the onset of the low-energy band. In fact, if the disorder is
sufficiently strong (and the clean band gap is not too large), the
gap can be closed by disorder broadening\cite{min-prb2011}.
Also, when the gap is present the
features of the band edge are qualitatively different between the BA and
the SCBA. Specifically, the divergence present in the clean DOS at the
band edge persists as a sharp spike in the BA, but is rounded off in the
SCBA, indicating that the DOS approaches zero in the low-density limit
for the SCBA but continues to diverge in the BA. 
Probes which are sensitive to the DOS, such as the tunneling conductance
in atomic force microscopy, capacitance measurements, or direct
measurement of $\frac{d\mu}{dn}$ via single electron transistor
spectroscopy should be sensitive to this difference at low carrier
density.
To highlight this difference, in the bottom row of Fig.~\ref{fig:DOS} we
show the low-energy conduction band edge for three different values of
the impurity density for each pair of self-energy and scattering
potential approximations. 
In the BA [Fig.~\ref{fig:DOS}(c,i)], increasing disorder decreases the
height of the peak at the band edge, and creates a dip in the region
where the band edge would be in the non-disordered case which widens
with increasing impurity density. 
The SCBA shows very different behavior at the band edge. The sharp peak
disappears, and the onset of the finite density of states occurs at a
slightly lower energy than in the BA. 
The CI and the SRI give similar results for the modification of the band
edge (except that the SRI is stronger for equivalent choice of $\nimp$
because of our choice of $V_0$) indicating that the SRI can even be used
to reliably approximate the low-density behavior of the DOS in the CI
case as long as $V_0$ is set correctly, as described in Appendix
\ref{app:SRCIcomp}.
The reason for this is that the screening wave vector is of the order of
the size of the Brillouin zone [see Fig~\ref{fig:diagrams}(d)] which is
large in comparison with the wave vectors near the K point which
dominate the contribution to the Green's function at low energy.
Therefore, the $q$ dependence in $V^C$ is negligable at this energy
scale and the interaction strength is almost independent of $q$, just as
it is in the short-range case.
However, for all impurity densities and for both approximations, the DOS
remains close to the non-disordered value away from the band edge with
the disorder broadening affecting the DOS mainly near the band edge.

\section{Optical conductivity} \label{sec:optcond}

\begin{figure*}[tbp]
	\centering
	\includegraphics[]{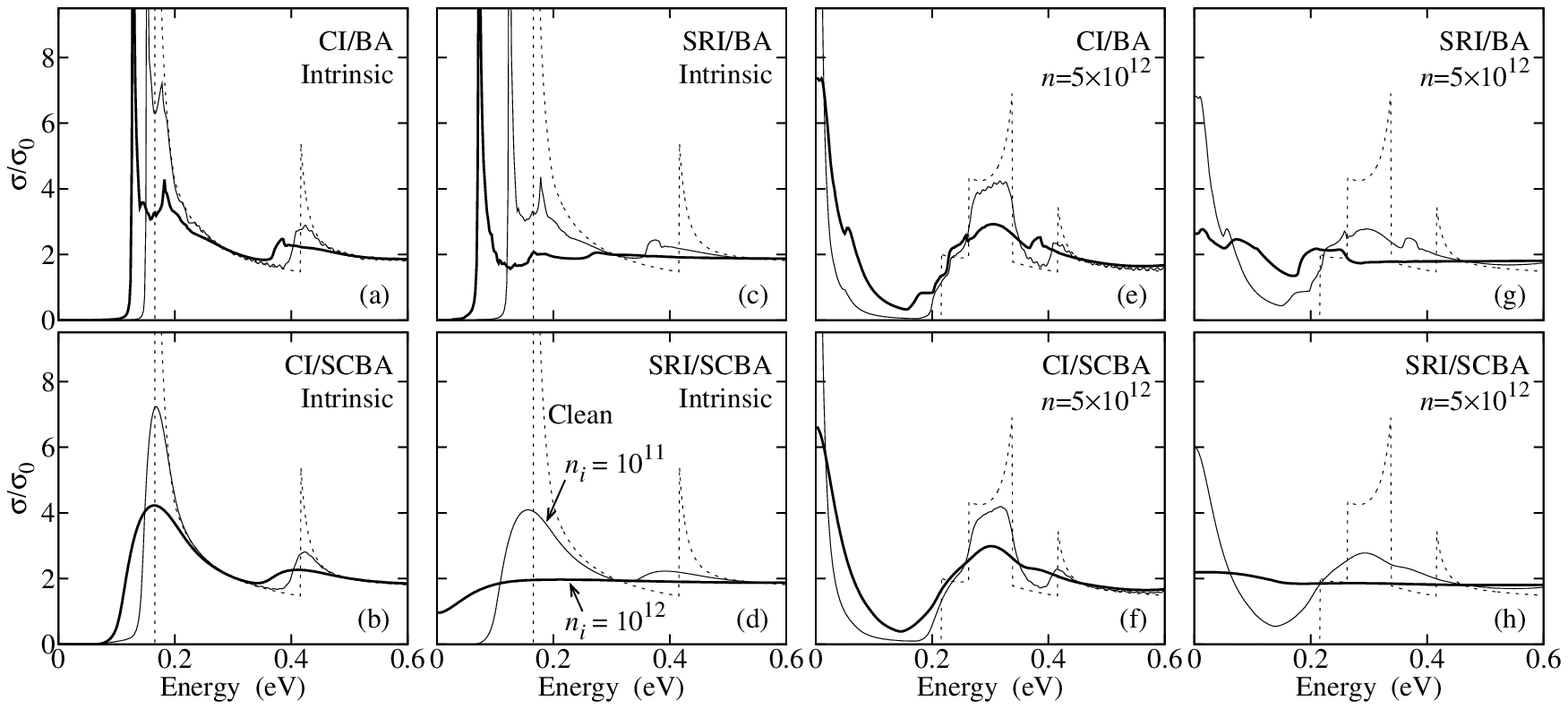}
	\includegraphics[]{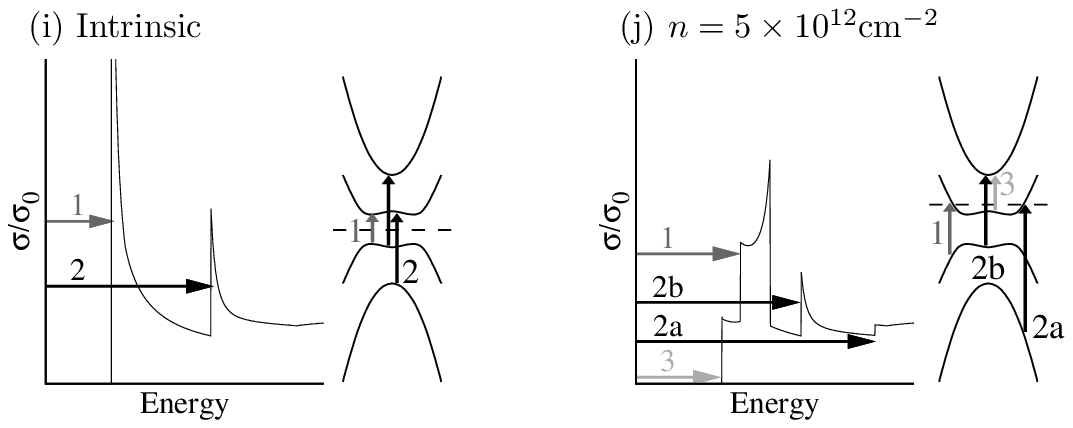}
	\caption{Panels (a--h) 
	The optical conductivity of bilayer graphene in units of
	$\sigma_0 = \frac{\pi e^2}{2 h}$ (the conductivity of monolayer
	graphene in the clean limit) with $u=200\meV$. 
	The type of scatterer (CI or SRI), approximation of the
	self-energy (BA or SCBA), and carrier density are labeled in each 
	panel. In all plots, the dotted line represents the contribution
	from the interband conductivity in the clean system (\textit{i.e.}
	the Drude peak is not included in the extrinsic case), the black
	line is $\nimp = 10^{12}\cmsq$, and the blue line is $\nimp = 5\times
	10^{12}\cmsq$. For the finite disorder cases, the intraband
	contribution is included. Throughout, we have assumed a SiO$_2$
	substrate with no top gate when defining the dielectric environment
	(see text for full definition of parameters). 
	Panels (i), (j) show the transitions which give rise to each feature
	in the non-disordered conductivity.
	\label{fig:optcond}}
\end{figure*}

We now turn our attention to the optical conductivity of gapped bilayer
graphene. This is an important quantity from the experimental
perspective since the optical identification of graphene
\cite{blake-apl2007,abergel-apl2007} and the optical determination of
the band gap\cite{li-prl2009,kuzmenko-prb2009,mak-prl2009} both depend
crucially on a thorough understanding of the optical conductivity. 
The optical conductivity in the absence of disorder was calculated some
time ago \cite{nicol-prb2008,min-prl2009}, but we extend this analysis
to include the effects of disorder using the same CI and SRI models
discussed before.
The optical conductivity is given by
\begin{multline}
	\sigma(E) = 2g_\mathrm{s} g_\mathrm{v} \frac{e^2}{h} 
	\sum_{\lambda,\lambda'} \int \frac{k' dk'}{2\pi} 
	\int^{\EF}_{\EF-E} \frac{dE'}{E} \\
	\times M^2_{\lambda\lambda'}(k') \mathrm{Im} G_{\vk'\lambda}(E')
	\mathrm{Im} G_{\vk'\lambda'}(E'+E)
	\label{eq:optconddef}
\end{multline}
where
\begin{equation}
	M^2_{\lambda\lambda'}(k) = \int_0^{2\pi} \frac{d\phi}{2\pi}
	\left| \left\langle \lambda,k,0| \hbar \hat{v}_x | \lambda', k, \phi
	\right\rangle \right|^2.
\end{equation}
The derivation of these equations is given in Appendix \ref{app:optcond}. 
We use $\eta=1\meV$ and show the results in Fig.~\ref{fig:optcond} for
the four combinations of scattering mechanism and approximation for the
self-energy.
Figure \ref{fig:optcond}(a)--(d) show the intrinsic case, and the origin
of each of the features in the curves is sketched in
Fig.~\ref{fig:optcond}(i).
In the CI shown in Fig.~\ref{fig:optcond}(a--b), the peak associated
with the interband transitions is shifted to lower energy with
increasing disorder. 
In the BA [Fig.~\ref{fig:optcond}(a)] the sharp spike at the band edge
persists even in the presence of strong disorder, while in the SCBA,
[Fig.~\ref{fig:optcond}(b)] it is rounded off and a significant tail
develops. 
Also, in the BA, there is a small peak near the energy
$E^0_{\mathrm{gap}}=166\meV$ which is an artifact of this approximation
and is not present in the SCBA.
These features directly reflect the related structure in the DOS. 
Comparing with the short-range interaction in
Fig.~\ref{fig:optcond}(c--d), it is clear that the qualitative behavior
of the optical conductivity near the band edge is identical to the CI.
However, because the strength of the interaction remains constant at
large wave vectors, the structure near the onset of the transitions
including the split band is very strongly modified even for small values
of the disorder.

We have taken the carrier density of $n=5\times 10^{12}\cmsq$ as an
example of the structure of the optical conductivity in the extrinsic
case. Figures~\ref{fig:optcond}(e--h) show the results for the same
parameters as for the calculations of intrinsic BLG. 
At this density, the Fermi energy is well above the sombrero region and
therefore the effects of this low energy structure are not present.
The features of the interband part of the optical conductivity are 
sketched in Fig.~\ref{fig:optcond}(j).
The additional structure in comparison to the intrinsic case is due to
the Fermi-blocking of some transitions, and the additional transitions
from the low-energy conduction band.
The large peak at low energy is the intraband contribution (called the
Drude peak), and is absent in the intrinsic case. 
The most obvious conclusion is that all the
structure that is due to the non-equal transition energies is very
quickly made indistinguishable by disorder, and even at
$\nimp=10^{12}\cmsq$, the structures associated with the different
interband transitions which include the split band become blurred into a
single peak. 
Finally, for strong enough disorder, the interband transitions become
blurred with the intraband peak making identification of the individual
features of the conductivity impossible.
Note that as the carrier density varies, the energies associated with
transitions 1 and 2a will change, strongly modifying the optical
conductivity. Therefore, accurate control of the density is essential
for extracting the correct optical conductivity.

\section{Band gaps} \label{sec:gaps}

\begin{figure}[tbp]
	\centering
	\includegraphics[]{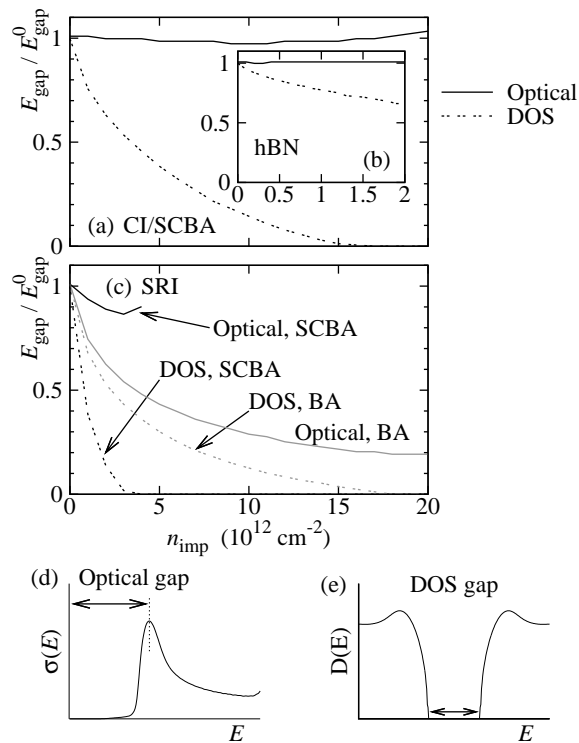}
	\caption{The band gaps extracted from the optical conductivity of
	intrinsic bilayer graphene and the DOS. (a) SCBA approximation for
	the CI on an SiO$_2$ substrate; (b) the same approximation but for a
	high quality hBN substrate (note the difference in the range of the
	horizontal axis); (c) SRI for BA and SCBA; (d), (e) definition of
	optical and DOS gaps.
	\label{fig:gaps}}
\end{figure}

We now describe the effect of disorder on the band gap as measured by
optical experiments in bilayer graphene and compare to the gap predicted
by the DOS calculations. 
There have been several attempts to determine the band gap via
measurement of the optical conductivity in absorption or reflection
measurements\cite{zhang-nature2009, li-prl2009, kuzmenko-prb2009, 
mak-prl2009}. 
Figure \ref{fig:optcond}(i) shows the transitions in the intrinsic case.
The transition between the low-energy valence and conduction bands is
labeled as transition 1. This transition gives a very clear peak in the
optical conductivity, and since there is no Drude peak in the intrinsic
case, it serves as a clear way to define the band gap. 
The onset of the peak occurs at $E^{(0)}_{\mathrm{gap}}$ in the clean
limit, and therefore serves as the definition of the gap in this case.
This definition is shown schematically in Fig.~\ref{fig:gaps}(d). 
Figure \ref{fig:optcond}(j) shows the transitions in the extrinsic case,
and we refer to this sketch in order to explain the procedure for
extracting the gap. The direct transition between the low-energy
conduction and valence bands [labeled transition 1 in
Fig.~\ref{fig:optcond}(j)] is problematic for
the extraction of the gap because the energy of the transition is
Fermi-blocked for the lowest energy excitations and therefore does not
directly give the band gap. Some experiments\cite{mak-prl2009} instead
computed the gap as the difference in energy between transition 2b and
transition 3, but this is also unreliable since even a small amount of
disorder masks the intricate structure of the optical conductivity.
Other experiments \cite{zhang-nature2009, kuzmenko-prb2009} fit single
particle theory with phenomenological disorder broadening to the
experimental data, and while this procedure is more reliable than
identifying particular peaks with individual transitions, it cannot take
into account the specifics of the impurity-electron interaction,
particularly at the band edge. For the purposes of the following
discussion, we define the energy of the optical band gap as the top of
the first peak in the optical conductivity in the intrinsic case, as
indicated in Fig.~\ref{fig:gaps}(d). Figure
\ref{fig:gaps}(a) shows that this point is very robust against
disorder for the CI in the SCBA [compare to the position of the top of the
peak in Fig.~\ref{fig:optcond}(b)]. The peak in the BA moves to lower
energy with disorder, implying that the measured optical gap in this
approximation is not robust compared with SCBA. Similarly, the
short-range interaction for the SCBA leads to such strong modification
of the peak structure that it is difficult to define the optical band
gap at significant impurity concentration. 
This information is shown in Fig.~\ref{fig:gaps}(a) where the optical
band gap is plotted as a function of impurity density. In the main
panel, a SiO$_2$ substrate is assumed. The solid lines denote the
optical gap, and we see that for the CI in the SCBA [solid black line in
Fig.~\ref{fig:gaps}(a)], the gap does not close. 
In contrast, for the SRI [solid black line in Fig.~\ref{fig:gaps}(b)] it
is only possible to define the band gap for a small range of impurity
concentration, and the SRI in the BA [solid gray line in
Fig.~\ref{fig:gaps}(b)] decreases rapidly.

We compare this to the DOS gap, which we define as the energy difference
between the onset of the DOS associated with the conduction and valence
bands as sketched in Fig.~\ref{fig:gaps}(e). This is plotted in
Fig.~\ref{fig:gaps}(a,b) using the dashed lines as labeled in each plot. 
It is clear that the effect of disorder is to rapidly shrink the size of
the DOS gap in all approximations. 
For reference, we point out that using very high quality substrates,
such as hexagonal boron nitride (h-BN) \cite{dean-natnano2010,
dassarma-prb2011},
where the impurity concentration is up to an order of magnitude smaller
than in SiO$_2$ substrates, the difference between the optical and DOS
gaps defined above will be significantly smaller. This is demonstrated
in the inset to Fig.~\ref{fig:gaps}(a), where the optical and DOS gaps
for bilayer graphene on h-BN are plotted for the CI in the SCBA, for a
reduced range of impurity density.

\section{Summary and conclusion} \label{sec:summary}

In summary, we have investigated the effect of scattering from charged
impurities and short-range potentials on the self-energy, density of
states, and optical conductivity in gapped bilayer graphene using both
the Born approximation and the self-consistent Born approximation. We
have demonstrated that these two approximations give qualitatively
different results for the DOS at the band edge, where the SCBA predicts
that the divergence of the DOS seen in the clean limit and in the BA
will be rounded off. Both approximations for the self-energy yield a DOS
where the gap between the conduction and valence bands closes with
increasing disorder strength. The main differences between the CI
and the SRI occur at the onset of the split bands. 
For our choices of parameters, the SRIs are stronger at these energies
and therefore the shift in the band is larger than for CI. 
We also introduced a self-consistent procedure to provide a
computationally advantageous approximation for the low-energy properties
in the CI via the SRI.
We then demonstrated that the optical conductivity in the presence of
disorder is a highly complicated function, the qualitative features of
which depend crucially on the disorder. Our two main conclusions are
that even a small level of disorder is sufficient to blur the main
features of the optical conductivity of extrinsic graphene, and this
makes the determination of the band gap via this measurement highly
problematic. In contrast, the peak associated with the interband
transition of the optical conductivity in the intrinsic case is rather
robust against the effects of disorder in the CI/SCBA case. Finally, we
discussed the evolution of the band gap in the DOS and extracted from
optical measurements of intrinsic bilayer graphene as a function of
impurity density, finding that the DOS gap is quickly reduced with
increasing disorder, and eventually will be closed completely. In
contrast, the gap extracted from the optical conductivity of intrinsic
bilayer graphene in the SCBA is much more robust against disorder.
This finding, that the DOS gap is always substantially smaller than the
nominal optical gap (see Fig.~\ref{fig:gaps}) defined operationally for
gapped BLG in the presence of disorder, may provide an explanation for
why the experimental transport gap extracted for gapped BLG appears to
be substantially smaller than both the theoretical single particle band
gap and the experimentally extracted optical gap \cite{zhang-nature2009,
kuzmenko-prb2009, li-prl2009, mak-prl2009, oostinga-natmat2007,
zou-prb2010, xiao-prb2010, yan-nl2010, xia-nl2010, taychatanapat-prl2010}. 
Our results clearly indicate that this discrepancy between transport and
optical gaps should disappear in systems with very little disorder such
as high mobility suspended graphene or high quality graphene on h-BN
substrates.
It is also interesting to note that our conclusion regaring the large
difference between transport and single particle (or optical) gaps is
completely consistent with a recent theoretical conclusion based on
percolation considerations of BLG graphene in the presence of
electron-hole puddles\cite{rossi-arXiv}.

We thank US-ONR and NRI-SWAN for support.

\appendix
\section{Approximation of Coulomb scattering by short-range scattering
\label{app:SRCIcomp}}

\begin{figure}[tbp]
	\centering
	\includegraphics[]{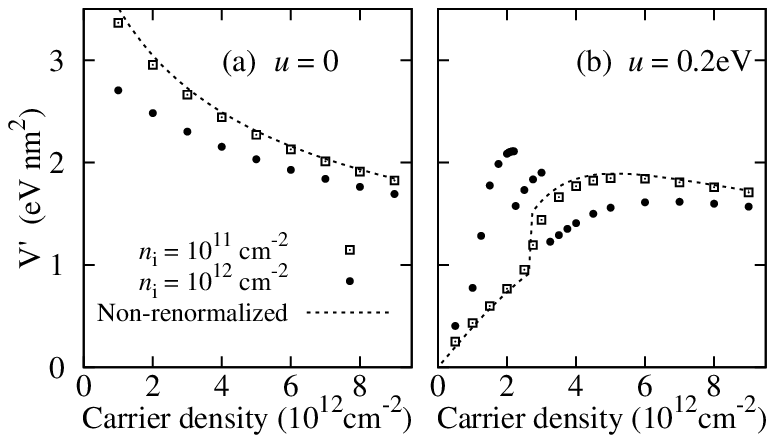}
	\includegraphics[]{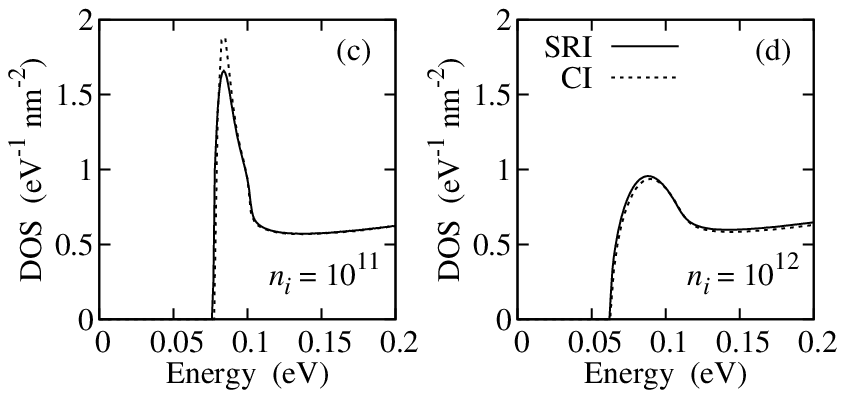}
	\caption{(a), (b) The SRI strength set by the effective CI, and
	renormalized to approximate the TF screening.
	(c), (d) The direct comparison of the DOS for CI and SRI for
	$u=0.2\mathrm{eV}$. The impurity density is labeled in each plot. In
	(c), $V'=2\mathrm{eV} \mathrm{nm}^2$ and in (d) $V'=1.6\mathrm{eV}
	\mathrm{nm}^2$.
	\label{fig:V0renorm}}
\end{figure}

In this appendix, we discuss a method for approximating the low-energy
properties of the CI using the SRI.
The physical mechanisms of the SRI and the CI are not related to each
other since different types of disorder in the environment of the
graphene will lead to each. However, as we have shown in this paper, 
the SRI and CI give qualitatively similar behavior for the physical
observables such as the optical conductivity.
If one is interested in transport characteristics, the compressibility
of the electron liquid, or other quantities that depend on the carrier
density, the CI is a much more time-consuming approximation to use
because the screening properties vary with the carrier density $n$ 
and hence the
self-energies also depend on it. This implies that the SCBA procedure
must be implemented for every value of $n$. This is not the case for the
SRI, where the impurity interaction is constant with $n$ and so one
self-energy function is valid for all carrier densities.
Therefore, it would be advantageous if we could describe a method for
approximating the CI by suitable choices of parameters for the SRI. 
In this appendix, we denote the value of $V_0$ set with reference to the
CI by the notation $V'$.
The obvious method for doing this is to set 
\begin{equation*}
	V' = V^C(0) = \frac{2\pi e^2}{\kappa \qs}
\end{equation*}
which is still a function of $n$ via the screening wave vector.
If this function is reasonably flat then a suitable constant value can
be chosen as an approximation and this step removes the dependence on
$n$ from the self-energy.
The dashed lines in Fig.~\ref{fig:V0renorm} (labeled `non-renormalized'
in the legend) show this for $u=0$ and
$u=0.2\mathrm{eV}$. We see that for higher carrier density ($n>2\times
10^{12}\cmsq$) there are no sharp changes in the value of $V'$
determined by this method, and that the slope is relatively modest.
However, at low carrier density in the gapped case, the interaction
strength becomes small and this approximation will not work.

A further sophistication can be introduced to account for the effects of
disorder on the density of states and hence on $\qs$.
In this approach, the disordered DOS is computed in the BA using the
value of $V'$ obtained above, and the renormalized Fermi energy $\EF'$
is extracted by
solving
\begin{equation*}
	\int_0^{E'_\mathrm{F}} D(E) dE = n
\end{equation*}
where we assume that the disorder does not change the carrier density.
This Fermi energy and DOS function can be used to set a new value of
$V'$, and this procedure is iterated until convergence is reached for
$V'$.
Doing this for two disorder concentrations in the gapped and gapless
cases are given in Figs.~\ref{fig:V0renorm}(a,b).
These plots show that selecting $V'\approx 2 \mathrm{eV} \mathrm{nm}^2$
will be a reasonable approximation for higher carrier density.
In Figs.~\ref{fig:V0renorm}(c,d) we show a comparison of the density of
states near the band edge in the SCBA for the SRI and CI for
$u=0.2\mathrm{eV}$ and two values of the impurity concentration. When
$\nimp = 10^{11}\cmsq$ we take $V'=2 \mathrm{eV}\mathrm{nm}^2$ and when
$\nimp = 10^{12}\cmsq$ we take $V'=1.6\mathrm{eV}\mathrm{nm}^2$. We see
that the two scattering mechanisms with these parameters give almost
identical results for the DOS.

\section{Optical conductivity} \label{app:optcond}

In this appendix, we present the details of the calculation of the
optical conductivity in the presence of disorder.
The real part of conductivity at a frequency $\omega$ can be calculated
from the Kubo formula \cite{mahan-2000} 
\begin{equation}
	\sigma(\omega)= {\rm Re} 
	\left[ \frac{i}{\omega} \Pi_{xx}^{R}(\omega)\right],
	\label{eq:kubo}
\end{equation}
where $\Pi_{\alpha\beta}^{R}(\omega)$ is the Fourier transform of the
real-time retarded current-current correlation function defined by
\begin{equation}
	i\hbar \Pi^{R}_{\alpha \beta}(\omega) = 
	\frac{1}{\Omega} \int_0^{\infty} dt \, e^{i\omega t}
	\left< \left[ \hat{J}_{\alpha}(t),\hat{J}_{\beta}(0) \right] \right>.
\end{equation}
Here $\Omega$ is the area of the system and $\hat{J}_{\alpha}$ is the
electric current operator defined by
\begin{equation}
	\hat{J}_{\alpha} = (-e) \sum_{\lambda,\lambda',\vk} 
	v_{\alpha}^{\lambda'\lambda}(\vk) c_{\vk\lambda'}^{\dagger} 
	c_{\vk\lambda},
\label{eq:current}
\end{equation}
where
$v_{\alpha}^{\lambda'\lambda}(\vk)=\left<\vk\lambda',\right|
v_{\alpha}\left|\vk,\lambda\right>$,
$c_{\vk\lambda}^{\dagger}$ and $c_{\vk\lambda}$ are creation and
annihilation operators for a state $\left|\lambda,\vk\right>$,
respectively.

A real-time retarded correlation function can be easily calculated from
a finite-temperature correlation function through the analytic
continuation in the complex $\omega-$space. The finite-temperature
current-current correlation function is defined by 
\begin{equation}
-\hbar\Pi_{\alpha\beta}(i\nu_n)=\frac{1}{\Omega}\int_0^{\beta\hbar}d\tau e^{i\nu_n\tau} \left<T_{\tau} J_{\alpha}(\tau) J_{\beta}(0)\right>,
\end{equation}
where $T_{\tau}$ is a time-ordering operator for an imaginary time
$\tau$, $\nu_n = \frac{2n\pi}{\beta\hbar}$ is a Matsubara frequency and
$\beta=1/(k_{\rm B}T)$.

Then the lowest-order correlation function is given by
\begin{multline}
\label{eq:correlation_T}
\hbar\Pi_{\alpha\beta}(i\nu_n) = \frac{e^2}{\beta\hbar V}
\sum_{\lambda,\lambda',\vk} v_{\alpha}^{\lambda'\lambda}(\vk)
v_{\beta}^{\lambda\lambda'}(\vk) \\
\times g_{\vk\lambda}(i\omega_n)  g_{\vk\lambda'}(i\omega_n+i\nu_n),
\end{multline}
where $g_{\vk\lambda}(i\omega_n)$ is the finite-temperature Green's
function defined by the Fourier transform of
$g_{\vk\lambda}(\tau)=-\left<T_{\tau} c_{\vk\lambda}(\tau)
c_{\vk\lambda}^{\dagger} (0)\right>$. Note that the spectral
representation of $g_{\vk\lambda}(i\omega_n)$ is given by
\begin{equation}
\label{eq:green_T}
g_{\vk\lambda}(i \omega_n)=\int \frac{d\omega'}{2\pi}
\frac{\rho_{\lambda}(\vk,\omega')} {i\omega_n-\omega'},
\end{equation}
where $\rho_{\lambda}(\vk,\omega)$ is a spectral weight function with
$\rho_{\lambda}(\vk,\omega)=-2{\rm
Im}\left[g_{\vk\lambda}(\omega+i\eta)\right]$.
From Eqs.~\eqref{eq:correlation_T} and \eqref{eq:green_T}, 
\begin{multline}
	\label{eq:correlation_T_lowest}
	\hbar\Pi^{}_{\alpha\beta}(i\nu_n) = -\frac{1}{V}
		\sum_{\vk,\lambda,\lambda'}
		\int \frac{d\omega}{2\pi} \frac{d\omega''}{2\pi} 
		\rho_{\lambda}(\vk,\omega') \rho_{\lambda'}(\vk,\omega'') \\
	\times v_{\alpha}^{\lambda'\lambda}(\vk) 
		v_{\beta}^{\lambda\lambda'}(\vk) 
		\frac{f(\omega'')-f(\omega')}{i\nu_n-(\omega''-\omega')},
\end{multline}
where $f(\omega)=\left[e^{\beta(\hbar\omega-\mu)}+1\right]^{-1}$ and
$\mu$ is the chemical potential. Here the following frequency sums for
fermions at finite temperatures were used: 
\begin{eqnarray}
&&\frac{1}{\beta\hbar} \sum_{\omega_n} \frac{e^{i\omega_n\eta}}{i\omega_n-\omega}=f(\omega), \\
&&\frac{1}{\beta\hbar} \sum_{\omega_n}
\frac{e^{i\omega_n\eta}}{(i\omega_n-\omega')(i\omega_n+i\nu_n-\omega'')}=-\frac{f(\omega'')-f(\omega')}{i\nu_n-(\omega''-\omega')}. \nonumber
\end{eqnarray}

Through the analytic continuation of $i\nu_n\rightarrow \omega+i\eta$,
the conductivity expression can be derived from Eqs.~\eqref{eq:kubo} and
\eqref{eq:correlation_T_lowest}. The energy bands have rotational symmetry
and by taking this into account along with the spin and valley degeneracy,
we finally get the conductivity expression at zero-temperature as in
Eq.~\eqref{eq:optconddef}.

\end{document}